\documentclass[10pt]{revtex4}
\usepackage{latexsym}
\usepackage{epsfig}
\usepackage{amsmath}

\begin{document}

\author{M. H. Dehghani$^{1,2}$ \footnote{mhd@shirazu.ac.ir}, A. Sheykhi $^{1,2}$\footnote{asheykhi@shirazu.ac.ir} and R. Dehghani $^{1}$}
\title{Thermodynamics of Quasi-Topological Cosmology}
\address{$^1$  Physics Department and Biruni Observatory, College of
Sciences, Shiraz University, Shiraz 71454, Iran\\
         $^2$  Research Institute for Astronomy and Astrophysics of Maragha (RIAAM), P. O. Box 55134-441, Maragha, Iran}
         
\begin{abstract}
In this paper, we study thermodynamical properties of the apparent horizon
in a universe governed by quasi-topological gravity. Our aim is twofold.
First, by using the variational method we derive the general form of
Friedmann equation in quasi-topological gravity. Then, by applying the first
law of thermodynamics on the apparent horizon, after using the entropy
expression associated with the black hole horizon in quasi-topological
gravity, and replacing the horizon radius, $r_{+}$, with the apparent
horizon radius, $\tilde{r}_{A}$, we derive the corresponding Friedmann
equation in quasi-topological gravity. We find that these two different
approaches yield the same result which show the profound connection between
the first law of thermodynamics and the gravitational field equations of
quasi-topological gravity. We also study the validity of the generalized
second law of thermodynamics in quasi-topological cosmology. We find that,
with the assumption of the local equilibrium hypothesis, the generalized
second law of thermodynamics is fulfilled for the universe enveloped by the
apparent horizon for the late time cosmology.
\end{abstract}

\maketitle

\section{Introduction\label{Intr}}

The most general Lagrangian which keeps the field equations of motion for
the metric of second order, as the pure Einstein-Hilbert action, is Lovelock
Lagrangian \cite{Lov}. This Lagrangian is constructed from the dimensionally
extended Euler densities and can be written as
\begin{equation}  \label{LL}
\mathcal{L}=\sum_{p=0}^{m}\alpha_{p}\mathcal{L}_{p},
\end{equation}
where $\alpha_{p}$ and $\mathcal{L}_{p}$ are arbitrary constant and Euler
density, respectively. In an $(n+1)$-dimensional spacetime $m=[n/2]$. $%
\mathcal{L}_{0}$ set to be one, and therefore $\alpha_{0}$ plays the role of
the cosmological constant. Because of the topological origin of the Lovelock
terms, the second order (Gauss-Bonnet) term does not have any dynamical
effect in four dimensions. Similarly, the cubic interaction only contributes
to the equations of motion when the bulk dimension is seven or greater. In
other words, although the equations of motion of $p$th order Lovelock
gravity are second-order differential equations, the $p$th order Lovelock
term has no contribution to the field equations in $2p$ and lower
dimensions. Is it possible to construct a gravitational action with cubic
curvature interactions or higher which has contribution in five dimension?
The answer is positive and the corresponding theory is called
``quasi-topological" gravity which was recently proposed in Refs. \cite%
{Olive1,Myer1} and \cite{MHD2} with cubic and quartic terms of Riemann
tensor, respectively. This new gravitational theory provides a useful toy
model to study a broader class of four (and higher) dimensional CFT's,
involving three or more independent parameters \cite{Myer2}. Various aspects
of $p$th- order quasi-topological terms which have at most second order
derivatives of the metric in the field equations for spherically symmetric
spacetimes in five and higher dimensions except $2p$ dimensions have been
investigated \cite{MHD1,MHD3,MHD4,QT}.

Nowadays, it is a general belief that there is a profound connection between
the gravitational field equations and the laws of thermodynamics. It was
shown that the gravitational field equation of a static spherically
symmetric spacetime in Einstein, Gauss-Bonnet and more general Lovelock
gravity can be recast as the first law of thermodynamics \cite{Par}. The
studies were also extended to other gravity theories such as $f(R)$ gravity
\cite{Cai0} and scalar-tensor gravity \cite{Cai1}. In the cosmological
setup, it was shown that the differential form of the Friedmann equation of
Friedmann-Robertson-Walker (FRW) universe can be transformed to the first
law of thermodynamics on the apparent horizon \cite{Cai2,Cai3}. In the
context of brane cosmology, it was shown that the Friedmann equations on the
brane can be expressed as $dE=TdS+WdV$ on the apparent horizon \cite%
{Cai4,Shey1,Shey2}. This procedure also leads to extract an expression for
the entropy at the apparent horizon on the brane, which is useful in
studying the thermodynamical properties of the black hole horizon on the
brane \cite{Cai4,Shey1,Shey2}.

Is the inverse procedure also possible? That is starting from the first law
of thermodynamics to extract the general field equations of gravitational
theory. Jacobson \cite{Jac} was the first who disclosed that the Einstein
field equation can be derived from the relation between the horizon area and
entropy, together with the Clausius relation $\delta Q=T\delta S$. Also, in
the cosmological setup, it was shown that the corresponding Friedmann
equations of Einstein, Gauss-Bonnet and Lovelock gravity can be derived by
applying the energy balance relation $-dE=TdS$ to the apparent horizon of a
Friedmann- Robertson-Walker universe (FRW) with any spatial curvature \cite%
{CaiKim}. Here, $-dE$ is actually just the heat flux $\delta Q$ in \cite{Jac}
crossing the apparent horizon within an infinitesimal internal of time $dt$.
In the framework of Horava-Lifshitz gravity, it was shown that the
corresponding Friedmann equation cannot be derived by applying the first law
of thermodynamics on the apparent horizon and using the entropy expression
for static spherically symmetric black holes in this gravity theory \cite%
{ShHL}. The reason of failure seems to be due to the fact that
Horava-Lifshiz gravity is not diffeomorphism invariant \cite{Pad2}. Indeed,
the action of Horava-Lifshiz gravity is invariant only under a restricted
class of diffeomorphism \cite{Kiri}. This implies that the connection
between first law of thermodynamics and gravitational field equations is not
a generic feature of any theory of gravity.

In this paper we will address the question on the connection between
thermodynamics and gravity by investigating whether and how the relation can
be found in quasi-topological cosmology. This is the first study on the
quasi-topological cosmology and in particular investigating thermodynamical
aspects of this gravity theory. For this purpose, we first derive the
Friedmann equations in quartic and higher order quasi-topological gravity by
varying the action of the quasi-topological gravity. Then, to show the
consistency of this theory with thermodynamics, we extract the corresponding
Friedmann equations by applying the first law of thermodynamics, $%
dE=T_hdS_h+WdV$, on the apparent horizon of a FRW universe governed by
quasi-topological gravity. Our strategy is to pick up the entropy expression
associated with the black hole horizon in quasi-topological gravity and
replacing the black hole horizon radius $r_{+}$ by the apparent horizon
radius $\tilde{r}_{A}$. We will also examine the time evolution of the total
entropy, including the entropy associated with the apparent horizon in
quasi-topological gravity together with the matter field entropy inside the
apparent horizon. We find that, in the late time, the generalized second law
(GSL) of thermodynamics is fulfilled for the universe governed by
quasi-topological gravity.

This paper is outlined as follows. In the next section, we introduce the
action of the quasi-topological gravity and derive the general form of the
Friedmann equation by using the variational method in this gravity theory.
In section III, we extract the Friedmann equation of quartic
quasi-topological cosmology by applying the first law of thermodynamics, $%
dE=T_hdS_h+WdV$, on the apparent horizon. We also generalize our study to
higher order quasi-topological theory in this section. We investigate the
validity of GSL of thermodynamics for a universe enveloped by the apparent
horizon in quasi-topological gravity in section IV. We finish our paper with
conclusions in section V.

\section{Quasi-Topological Cosmology}

\label{quasi111} In this section we derive the field equations governing the
evolution of the universe in quasi-topological gravity. The most general
gravitational theory which produces second-order equation of motion is the $%
i $-order Lovelock gravity with action \cite{Lov}
\begin{equation}
I_{\mathrm{G}}=\frac{1}{16\pi G_{n+1}}\int d^{n+1}x\sqrt{-g}\left(-2\Lambda
+\sum_{i=0}^{m}\alpha_{i}\mathcal{L}_{i}+\mathcal{L}_{\mathrm{M}}\right),
\label{Act1}
\end{equation}
where $\Lambda$ is the cosmological constant, and the $\alpha_{i}$'s are
Lovelock coefficients with dimensions $(\mathrm{length})^{2i-2}$, and $%
\mathcal{L}_i$ is the $i$th order Lovelock Lagrangian
\begin{equation}
\mathcal{L}_{i}=\frac{1}{2^{i}}\delta _{\nu _{1}\text{ }\nu _{2}\cdots \nu
_{2i}}^{\mu _{1}\mu _{2}\cdots \mu _{2i}}R_{\mu _{1}\mu _{2}}^{%
\phantom{\mu_1\mu_2}{\nu_1\nu_2}}\cdots R_{\mu _{2i-1}\mu _{2i}}^{%
\phantom{\mu_{2i-1} \mu_{2i}}{\nu_{2i-1} \nu_{2i}}}.  \label{LoveLag}
\end{equation}
In the action (\ref{Act1}), the term proportional to $\alpha_i$ contributes
to the equations of motion in dimensions with $n \geq 2i$. For example, the
terms associated to $i=3$ or higher do not contribute to the field equations
in five dimensions. Recently, a new gravity theory called quasi-topological
gravity has been introduced, which has contribution to the field equations
in five dimensions from the $i$-order ($i\geq3$) term in Riemann tensor.

The gravity part of the action of the quartic quasi-topological theory in $%
(n+1)$-dimensions in the absence of cosmological constant is given by \cite%
{MHD2}
\begin{equation}
I=\int d^{n+1}x\left( \mathcal{L}_{G}+\mathcal{L}_{M}\right) ,  \label{Act2}
\end{equation}%
where $\mathcal{L}_{M}$ is the Lagrangian of the matter and
\begin{equation}
\mathcal{L}_{G}=\frac{\sqrt{-g}}{16\pi G_{n+1}}\left( {\mu }_{1}\mathcal{L}%
_{1}+{\mu }_{2}\mathcal{L}_{2}+{\mu }_{3}\mathcal{X}_{3}+{\mu }_{4}\mathcal{X%
}_{4}\right) .  \label{act02}
\end{equation}%
In Eq.(\ref{act02}) $\mathcal{L}_{1}=R$ is the Einstein-Hilbert Lagrangian, $%
\mathcal{L}_{2}=R_{abcd}{R}^{abcd}-4{R}_{ab}{R}^{ab}+{R}^{2}$ is the second
order Lovelock (Gauss-Bonnet) Lagrangian, $\mathcal{X}_{3}$\ is the
curvature-cubed Lagrangian \cite{Myer1}
\begin{eqnarray}
\mathcal{X}_{3} &=&R_{ab}^{cd}R_{cd}^{\,\,e\,\,\,f}R_{e\,\,f}^{\,\,a\,\,\,b}+%
\frac{1}{(2n-1)(n-3)}\left( \frac{3(3n-5)}{8}R_{abcd}R^{abcd}R\right.
\notag \\
&&-3(n-1)R_{abcd}R^{abc}{}_{e}R^{de}+3(n+1)R_{abcd}R^{ac}R^{bd}  \notag \\
&&\left. +\,6(n-1)R_{a}{}^{b}R_{b}{}^{c}R_{c}{}^{a}-\frac{3(3n-1)}{2}%
R_{a}^{\,\,b}R_{b}^{\,\,a}R+\frac{3(n+1)}{8}R^{3}\right) .  \label{X3}
\end{eqnarray}%
and $\mathcal{X}_{4}$ is the fourth order term of quasi-topological gravity
\cite{MHD2}
\begin{eqnarray}
\mathcal{X}_{4}\hspace{-0.2cm} &=&\hspace{-0.2cm}c_{1}R_{abcd}R^{cdef}R_{%
\phantom{hg}{ef}%
}^{hg}R_{hg}{}^{ab}+c_{2}R_{abcd}R^{abcd}R_{ef}R^{ef}+c_{3}RR_{ab}R^{ac}R_{c}{}^{b}+c_{4}(R_{abcd}R^{abcd})^{2}
\notag \\
&&\hspace{-0.1cm}%
+c_{5}R_{ab}R^{ac}R_{cd}R^{db}+c_{6}RR_{abcd}R^{ac}R^{db}+c_{7}R_{abcd}R^{ac}R^{be}R_{%
\phantom{d}{e}}^{d}+c_{8}R_{abcd}R^{acef}R_{\phantom{b}{e}}^{b}R_{%
\phantom{d}{f}}^{d}  \notag \\
&&\hspace{-0.1cm}%
+c_{9}R_{abcd}R^{ac}R_{ef}R^{bedf}+c_{10}R^{4}+c_{11}R^{2}R_{abcd}R^{abcd}+c_{12}R^{2}R_{ab}R^{ab}
\notag \\
&&\hspace{-0.1cm}%
+c_{13}R_{abcd}R^{abef}R_{ef}{}_{g}^{c}R^{dg}+c_{14}R_{abcd}R^{aecf}R_{gehf}R^{gbhd},
\label{X4}
\end{eqnarray}%
where the coefficients $c_{i}$ are given by
\begin{eqnarray*}
c_{1} &=&-\left( n-1\right) \left( {n}^{7}-3\,{n}^{6}-29\,{n}^{5}+170\,{n}%
^{4}-349\,{n}^{3}+348\,{n}^{2}-180\,n+36\right) , \\
c_{2} &=&-4\,\left( n-3\right) \left( 2\,{n}^{6}-20\,{n}^{5}+65\,{n}^{4}-81\,%
{n}^{3}+13\,{n}^{2}+45\,n-18\right) , \\
c_{3} &=&-64\,\left( n-1\right) \left( 3\,{n}^{2}-8\,n+3\right) \left( {n}%
^{2}-3\,n+3\right) , \\
c_{4} &=&-{(n}^{8}-6\,{n}^{7}+12\,{n}^{6}-22\,{n}^{5}+114\,{n}^{4}-345\,{n}%
^{3}+468\,{n}^{2}-270\,n+54), \\
c_{5} &=&16\,\left( n-1\right) \left( 10\,{n}^{4}-51\,{n}^{3}+93\,{n}%
^{2}-72\,n+18\right) , \\
c_{6} &=&-32\,\left( n-1\right) ^{2}\left( n-3\right) ^{2}\left( 3\,{n}%
^{2}-8\,n+3\right) , \\
c_{7} &=&64\,\left( n-2\right) \left( n-1\right) ^{2}\left( 4\,{n}^{3}-18\,{n%
}^{2}+27\,n-9\right) , \\
c_{8} &=&-96\,\left( n-1\right) \left( n-2\right) \left( 2\,{n}^{4}-7\,{n}%
^{3}+4\,{n}^{2}+6\,n-3\right) , \\
c_{9} &=&16\left( n-1\right) ^{3}\left( 2\,{n}^{4}-26\,{n}^{3}+93\,{n}%
^{2}-117\,n+36\right) , \\
c_{10} &=&{n}^{5}-31\,{n}^{4}+168\,{n}^{3}-360\,{n}^{2}+330\,n-90, \\
c_{11} &=&2\,(6\,{n}^{6}-67\,{n}^{5}+311\,{n}^{4}-742\,{n}^{3}+936\,{n}%
^{2}-576\,n+126), \\
c_{12} &=&8\,{(}7\,{n}^{5}-47\,{n}^{4}+121\,{n}^{3}-141\,{n}^{2}+63\,n-9), \\
c_{13} &=&16\,n\left( n-1\right) \left( n-2\right) \left( n-3\right) \left(
3\,{n}^{2}-8\,n+3\right) , \\
c_{14} &=&8\,\left( n-1\right) \left( {n}^{7}-4\,{n}^{6}-15\,{n}^{5}+122\,{n}%
^{4}-287\,{n}^{3}+297\,{n}^{2}-126\,n+18\right) .
\end{eqnarray*}%
The action (\ref{Act2}) not only works in five dimensions, but also yields
second-order equations of motion for spherically symmetric spacetimes \cite%
{MHD2}.

Our aim here is to derive the corresponding Friedmann equations of quartic
and higher order quasi-topological gravity. We consider a homogeneous and
isotropic FRW universe in $(n+1)$-dimensions which is described by the line
element
\begin{equation}  \label{met1}
ds^2=-N(t) dt^2+a^2(t)\left[\frac{dr^2}{1-k r^2}+r^2 d\Omega^2_{n-1}\right],
\end{equation}
where $k$ is the spatial curvature constant with values $1,0$ and $-1$
correspond to closed, flat and open universe, respectively, and $%
d\Omega^2_{n-1}$ represents the line elements of an $(n-1)$-dimensional unit
sphere.

We use the variational method for deriving the Friedmann equation from the
action principle. Using metric (\ref{met1}), the Lagrangian of the fourth
order quasi-topological gravity can be written as
\begin{eqnarray}  \label{sgr122}
\mathcal{L}_G=\frac{n(n-1)}{16 \pi G_{n+1}} \frac{\sqrt{-\gamma}a^n} {%
N^{(n+3)/2}a^8}\left(-N^4 b_1+N^3 b_2\dot{a}^2+\frac{1}{3}N^2 b_3 \dot{a}^4+%
\frac{1}{5}N b_4\dot{a}^6+\frac{1}{7}b_{5}\dot{a}^8\right).  \label{L0}
\end{eqnarray}
where $\gamma$ is the determinant of the metric of $t$-constant
hypersurface, $b_i$'s are
\begin{eqnarray*}
b_1 &=&a^6 k+a^4 k^2 \hat{\mu}_2 l^2 +a^2 k^3\hat{\mu}_3 l^4+ k^4 \hat{\mu}%
_4 l^6 , \\
b_2 &=& a^6+2 a^4 k\hat{\mu}_2 l^2+3 a^2 k^2 \hat{\mu}_3 l^4+4 k^3 \hat{\mu}%
_4 l^6, \\
b_3 &=& a^4 \hat{\mu}_2 l^2+3 a^2 k \hat{\mu}_3 l^4+6 k^2 \hat{\mu}_4 l^6, \\
b_4 &=& a^2 \hat{\mu}_3 l^4+4 k \hat{\mu}_4 l^6, \\
b_5 &=& \hat{\mu}_4 l^6.
\end{eqnarray*}
and the dimensionless parameters $\hat{\mu}_{j}$'s are
\begin{eqnarray*}
&& \hat{\mu}_{1}=1 \text{ \ \ },\hat{\mu}_{2}=\frac{(n-2)(n-3)}{l^2}\mu_{2}
\text{ \ \ },\hat{\mu}_{3}=\frac{(n-2)(n-5)(3 n^2-9n+4)}{8 (2n-1) l^4}\mu_3
\text{ \ \ }, \\
&& \hat{\mu}_{4}=\frac{n(n-1)(n-2)^2(n-3)(n-7)(n^5-15 n^4+72 n^3-156 n^2+150
n-42)}{l^6}\mu_{4}.
\end{eqnarray*}

Varying the Lagrangian (\ref{L0}) with respect to $N(t)$, we arrive at
\begin{equation}
\frac{1}{\sqrt{-g}}\frac{\delta \mathcal{L}_{G}}{\delta N}=-\frac{n(n-1)}{%
32\pi G_{n+1}}\frac{1}{N^{(n+6)/2}a^{8}}\left[ N^{4}b_{1}+N^{3}b_{2}\dot{a}%
^{2}+N^{2}b_{3}\dot{a}^{4}+Nb_{4}\dot{a}^{6}+b_{5}\dot{a}^{8}\right]
\label{varn}
\end{equation}
On the other hand the variation of the Lagrangian of matter with respect to $%
g_{00}=-N$ leads to
\begin{equation}
\frac{1}{\sqrt{-g}}\frac{\delta \mathcal{L}_{M}}{\delta N}\equiv \frac{%
T_{0}^{0}}{2N}=\frac{\rho }{2N}.  \label{varnM}
\end{equation}
Now, one can absorb $N(t)$ in $t$ coordinate. That is, one can set $N(t)=1$.
Using (\ref{varn}) and (\ref{varnM}), one obtains
\begin{equation}
\sum_{i=1}^{4}\hat{\mu}_{i}l^{2i-2}\left( H^{2}+\frac{k}{a^{2}}\right) ^{i}=%
\frac{16\pi G_{n+1}}{n(n-1)}\rho .  \label{Fri2}
\end{equation}
As in the case of black hole solutions presented in \cite{MHD2}, the form of
the field equation (\ref{Fri2}) allows us to generalize this equation to the
case of $m$th-order quasi-topological gravity:
\begin{equation}
\sum_{i=1}^{m}\hat{\mu}_{i}l^{2i-2}\left( H^{2}+\frac{k}{a^{2}}\right) ^{i}=%
\frac{16\pi G_{n+1}}{n(n-1)}\rho .
\end{equation}

Here, we pause to study the field equations under a small perturbation
around the Friedmann-Robertson-Walker metric. The authors of \cite{Myer1},
examined the linearized equations of motion for a graviton perturbation
around the AdS metric in cubic quasi-topological gravity, and showed that
the linearized graviton equation in an AdS background is only a second-order
equation. Here, we examine the same fact under a small perturbation around
the FRW metri and find that the linearized field equation is a second order
equation. This fact is different from the small perturbation around FRW
metric in the new massive garvity \cite{Sinha}. In the latter case, the
linearized field equation contains more than two-derivative and one may have
ghosty vacuum at initial times while it becomes free of ghosts at later
times in the cosmological scenario \cite{Sinha}.

\section{Friedman Equation from the first law \label{FIRST}}

In this section we would like to derive the Friedmann equation of
quasi-topological cosmology by applying the first law of thermodynamics on
the apparent horizon of FRW universe. The entropy associated with the event
horizon of higher dimensional static spherically symmetric black holes in
cubic quasi-topological gravity has the following form \cite{Myer1,MHD1}
\begin{equation}  \label{Sqt1}
S_h=\frac{A}{4G_{n+1}} \left[1+\frac{2(n-1)}{n-3} \frac{{\hat{\mu}}_{2}l^2}{%
r_{+}^2} +\frac{3(n-1)}{n-5} \frac{{\hat{\mu}}_{3}l^4}{r_{+}^4}\right],
\end{equation}
where $n\neq 5$ and $G_{n+1}$ is the $(n+1)$ dimensional gravitational
constant. Here $A=n\Omega_{n} r_{+}^{n-1} $ is the surface area of the black
hole horizon, and
\begin{equation}
\Omega_{n}=\frac{\pi^{n/2}}{\Gamma(\frac{n+2}{2})}, \ \ \ \ \Gamma\left(%
\frac{n+2}{2}\right)=\left(\frac{n}{2}\right) \left(\frac{n-2}{2}\right)!.
\end{equation}
We further assume the entropy expression (\ref{Sqt1}) is also valid for the
apparent horizon of the FRW universe in quasi-topological gravity. Replacing
the horizon radius $r_+$ with the apparent horizon radius $\tilde{r}_{A}$,
the entropy expression (\ref{Sqt}) can be written
\begin{equation}  \label{Sqt2}
S_h=\frac{n\Omega_{n}}{4G_{n+1}}\tilde{r}_{A}^{n-1} \left[1+ \frac{\beta}{%
\tilde{r}_{A}^2} +\frac{\gamma}{\tilde{r}_{A}^4}\right],
\end{equation}
where we have defined
\begin{eqnarray}  \label{dS1}
\beta =\frac{2(n-1)}{n-3}\hat{\mu}_{2}l^2, \ \ \ \gamma =\frac{3(n-1)}{n-5}%
\hat{\mu}_{3}l^4.
\end{eqnarray}
Taking differential form of relation (\ref{Sqt2}), we have
\begin{eqnarray}  \label{dS1}
&&dS=\frac{\partial S}{\partial \tilde{r}_{A} }d \tilde{r}_{A} =\frac{%
n\Omega_{n}}{4G_{n+1}} \left[(n-1)\tilde{r}_{A}^{n-2}+\beta(n-3)\tilde{r}%
_{A}^{n-4}+\gamma(n-5)\tilde{r}_{A}^{n-6}\right]d\tilde{r}_{A}.  \notag
\end{eqnarray}
We rewrite the line element of the FRW metric as
\begin{equation}
ds^2={h}_{a b}dx^{a} dx^{b}+\tilde{r}^2d\Omega_{n-1}^2,
\end{equation}
where $x^0=t, x^1=r$, $\tilde{r}=a(t)r$, and $h_{ab}$=diag $(-1,
a^2/(1-kr^2))$ represents the two dimensional metric. The dynamical apparent
horizon, a marginally trapped surface with vanishing expansion, is
determined by the relation $h^{ab}\partial_{a}\tilde {r}\partial_{b}\tilde {r%
}=0$. It is a matter of calculation to show that the radius of the apparent
horizon for the FRW universe becomes \cite{Hay2}
\begin{equation}  \label{radius}
\tilde{r}_A=\frac{1}{\sqrt{H^2+k/a^2}}.
\end{equation}
The temperature associated with the apparent horizon is defined as $T_h =
\kappa/2\pi$, where $\kappa =\frac{1}{2\sqrt{-h}}\partial_{a}\left(\sqrt{-h}%
h^{ab}\partial_{b}\tilde {r}\right)$ is the surface gravity. It is easy to
show that the surface gravity at the apparent horizon of FRW universe can be
written as
\begin{equation}  \label{surgrav}
\kappa=-\frac{1}{\tilde r_A}\left(1-\frac{\dot {\tilde r}_A}{2H\tilde r_A}%
\right).
\end{equation}
Since for $\dot {\tilde r}_A< 2H\tilde r_A$, we have $\kappa< 0$, which
leads to the negative temperature, thus one may, in general, define the
temperature on the apparent horizon as $T_h=|\kappa|/2\pi$. In addition,
since we associate with the apparent horizon a temperature, thus one may
expect that the apparent horizon have a kind of Hawking radiation just like
a black hole event horizon. This issue was previously addressed \cite{cao},
by showing the connection between temperature on the apparent horizon and
the Hawking radiation. This study gives more solid physical implication of
the temperature associated with the apparent horizon.

The energy conservation law $\nabla_{\mu}T^{\mu \nu}=0$ leads to the
continuity equation in the form
\begin{equation}  \label{Cont}
\dot{\rho}+n H(\rho+p)=0,
\end{equation}
The next quantity we need to have is the work density. In our case it can be
calculated as \cite{Hay2}
\begin{equation}  \label{Work}
W=-\frac{1}{2} T^{\mu\nu}h_{\mu\nu}=\frac{1}{2}(\rho-p).
\end{equation}
The work density is regarded as the work done when the apparent horizon
radius changes from ${\tilde r}_A$ to ${\tilde r}_A+d{\tilde r}_A$. Then, we
suppose the first law of thermodynamics on the apparent horizon of the
universe in quasi-topological gravity holds and has the form
\begin{equation}  \label{FL}
dE = T_h dS_h + WdV,
\end{equation}
where $S_{h}$ is the entropy associated with the apparent horizon in
quasi-topological cosmology given in Eq. (\ref{Sqt2}). The term $WdV$ in the
first law comes from the fact that we have a volume change for the total
system enveloped by the apparent horizon. For a pure de Sitter space, $%
\rho=-p$, and the work term reduces to the standard $-pdV$, thus we obtain
exactly the standard first law of thermodynamics, $dE = TdS-pdV$.

Assuming the total energy content of the universe inside an $n $-sphere of
radius $\tilde{r}_{A}$ is $E=\rho V$, where $V=\Omega_{n}\tilde{r}_{A}^{n}$
is the volume enveloped by an $n$-dimensional sphere. Taking differential
form of the total energy, after using the continuity equation (\ref{Cont}),
we obtain
\begin{eqnarray}  \label{dE1}
dE =\rho n\Omega_{n} \tilde {r}_{A}^{n-1} d\tilde {r}_{A}+\Omega_{n}\tilde{r}%
_{A}^{n}\dot{\rho} dt =\rho n\Omega_{n} \tilde {r}_{A}^{n-1} d\tilde {r}_{A}
-n H \Omega_{n} \tilde{r}_{A}^{n}(\rho+p) dt.
\end{eqnarray}
Substituting Eqs. (\ref{dS1}), (\ref{Work}) and (\ref{dE1}) in the first law
(\ref{FL}) and using the the definition of the temperature associated with
the apparent horizon, we get the differential form of the Friedmann equation
in cubic quasi-topological gravity as
\begin{eqnarray}  \label{Fried1}
&&\frac{1}{8\pi G_{n+1} \tilde {r}_{A}}\left[\frac{n-1}{\tilde {r}_{A}^2}+
\frac{\beta(n-3)}{\tilde {r}_{A}^4}+ \frac{\gamma(n-5)}{\tilde {r}_{A}^6}%
\right] d\tilde {r}_{A} = H (\rho+p) dt.
\end{eqnarray}
Using the continuity equation (\ref{Cont}), we obtain
\begin{equation}  \label{Fried2}
\left[\frac{n-1}{\tilde {r}_{A}^3}+ \frac{\beta(n-3)}{\tilde {r}_{A}^5}+
\frac{\gamma(n-5)}{\tilde {r}_{A}^7}\right] d\tilde {r}_{A} = -\frac{8\pi
G_{n+1}}{n}d\rho.
\end{equation}
Integrating (\ref{Fried2}) yields
\begin{equation}  \label{Fried3}
\frac{1}{{\tilde {r}_{A}}^2}+\frac{\hat{\mu}_2l^2}{\tilde {r}_{A}^4}+\frac{%
\hat{\mu}_{3}l^4}{\tilde {r}_{A}^6}= \frac{16\pi G_{n+1}}{n(n-1)}\rho,
\end{equation}
where an integration constant has been absorbed into the energy density $%
\rho $. Substituting $\tilde {r}_{A}$ from Eq.(\ref{radius}) we obtain
\begin{eqnarray}  \label{Fried4}
&&H^2+\frac{k}{a^2}+\hat{\mu}_{2}l^2\left(H^2+\frac{k}{a^2}\right)^2 +\hat{%
\mu}_{3}l^4\left(H^2+\frac{k}{a^2}\right)^3 = \frac{16\pi G_{n+1}}{n(n-1)}%
\rho.
\end{eqnarray}
In this way we derived the $(n+1)$-dimensional Friedmann equation governing
the evolution of the universe in cubic order quasi-topological gravity by
applying the first law of thermodynamics on the apparent horizon.

The above analysis can be extended to higher order quasi-topological
gravity. The entropy associated with spherically symmetric black hole
solutions in quartic order of quasi-topological gravity is given by \cite%
{MHD2}
\begin{equation}  \label{Sqt}
S_h=\frac{A}{4G_{n+1}} \left[\hat{\mu}_1+\frac{2(n-1)}{(n-3)} \frac{\hat{\mu}%
_{2}l^2}{r_{+}^2}+\frac{3(n-1)}{(n-5)} \frac{\hat{\mu}_3l^4}{r_{+}^4}+\frac{%
4(n-1)}{(n-7)}\frac{\hat{\mu}_4l^6}{r_{+}^6}\right].
\end{equation}
The extension to higher order quasi-topological black holes are quite
straightforward and can be written in the compact form
\begin{equation}  \label{gens}
S_h=\frac{A}{4G_{n+1}}\sum^{m}_{i=1} i \frac{(n-1)}{(n+1-2i)}\frac{\hat{\mu}%
_{i}l^{2i-2}}{r^{2i-2}_{+}}.
\end{equation}
Using the same formalism as we have done in this section, we obtain the
general form of the Friedmann equation as
\begin{eqnarray}  \label{Fried11}
&&\hat{\mu}_1\left(H^2+\frac{k}{a^2}\right)+\hat{\mu}_2l^2\left(H^2+\frac{k}{%
a^2}\right)^2 +\hat{\mu}_3l^4\left(H^2+\frac{k}{a^2}\right)^3+\hat{\mu}_4l^6
\left(H^2+\frac{k}{a^2}\right)^4+... = \frac{16\pi G_{n+1}}{n(n-1)}\rho.
\end{eqnarray}
or in a compact form as
\begin{eqnarray}  \label{Fried122}
&&\sum^{\infty}_{i=1}\hat{\mu}_{i}l^{2i-2}\left(H^2+\frac{k}{a^2}\right)^i=
\frac{16\pi G_{n+1}}{n(n-1)}\rho.
\end{eqnarray}
Which is compatible with the field equation derived through the use of the
variation of the action.

\section{GSL in quasi-topological gravity\label{GSL}}

Next, we examine the time evolution of the total entropy. For this purpose,
we take the time derivative of Eq. (\ref{Fried11}) and get
\begin{equation}
\left[ \frac{1}{{\tilde{r}_{A}}^{3}}+\frac{2\hat{\mu}_{2}l^{2}}{\tilde{r}%
_{A}^{5}}+\frac{3\hat{\mu}_{3}l^{4}}{\tilde{r}_{A}^{7}}+\frac{4\hat{\mu}%
_{4}l^{6}}{\tilde{r}_{A}^{9}}+...\right] \dot{\tilde{r}}_{A}=-\frac{8\pi
G_{n+1}}{n(n-1)}\dot{\rho},  \label{GSL1}
\end{equation}
Using the continuity equation (\ref{Cont}) and solving for $\dot{\tilde{r}}%
_{A}$ we obtain
\begin{equation}
\dot{\tilde{r}}_{A}=\frac{8\pi G_{n+1}}{(n-1)}H(\rho +p)\left[ \frac{1}{{%
\tilde{r}_{A}}^{3}}+\frac{2\hat{\mu}_{2}l^{2}}{\tilde{r}_{A}^{5}}+\frac{3%
\hat{\mu}_{3}l^{4}}{\tilde{r}_{A}^{7}}+\frac{4\hat{\mu}_{4}l^{6}}{\tilde{r}%
_{A}^{9}}+...\right] ^{-1}.  \label{dotr1}
\end{equation}
Calculating $T_{h}\dot{S_{h}}$,
\begin{eqnarray}
T_{h}\dot{S_{h}} &=&\frac{1}{2\pi \tilde{r}_{A}}\left( 1-\frac{\dot{\tilde{r}%
}_{A}}{2H\tilde{r}_{A}}\right) \frac{n(n-1)\Omega _{n}}{4G_{n+1}}\tilde{r}%
_{A}^{n+1}  \notag  \label{TSh1} \\
&&\times \left[ \frac{1}{{\tilde{r}_{A}}^{3}}+\frac{2\hat{\mu}_{2}l^{2}}{%
\tilde{r}_{A}^{5}}+\frac{3\hat{\mu}_{3}l^{4}}{\tilde{r}_{A}^{7}}+\frac{4\hat{%
\mu}_{4}l^{6}}{\tilde{r}_{A}^{9}}+...\right] \dot{\tilde{r}}_{A},
\end{eqnarray}
and substituting $\dot{\tilde{r}}_{A}$ from Eq. (\ref{dotr1}) in it, we
obtain
\begin{equation}
T_{h}\dot{S_{h}}=n\Omega _{n}{\tilde{r}_{A}^{n}}H\rho (1+w)\left( 1-\frac{%
\dot{\tilde{r}}_{A}}{2H\tilde{r}_{A}}\right) ,  \label{TSh2}
\end{equation}
where we have defined the equation of state $w={p}/{\rho }$, as usual. Since
$T\geq 0$, we have $\dot{\tilde{r}}_{A}\leq 2H\tilde{r}_{A}$, thus the sign
of Eq. (\ref{TSh2}) depends on $w$. If $w\geq -1$, then $\dot{S_{h}}\geq 0$
and the second law of thermodynamics is fulfilled. However, some
astrophysical evidences show that our universe is currently accelerating and
in particular the equation of state parameter can cross the phantom line,
i.e. $w<-1$, indicating that the second law of thermodynamics, $\dot{S_{h}}%
\geq 0$, does not hold. Nevertheless, as we shall see below, the GSL of
thermodynamics, $\dot{S_{h}}+\dot{S_{m}}\geq 0$, is still preserved
throughout the history of the universe. In order to verify the GSL of
thermodynamics, we have to study the time evolution of the total entropy
including the entropy $S_{h}$ associated with the apparent horizon together
with the matter field entropy $S_{m}$ inside the apparent horizon. The
entropy of the universe inside the horizon can be related to its energy and
pressure by Gibbs equation \cite{Pavon2}
\begin{eqnarray}
T_{m}dS_{m}&=&d(\rho V)+pdV=Vd\rho +(\rho +p)dV  \notag  \label{Gib2} \\
&=&\Omega _{n}\tilde{r}_{A}^{n}d\rho +(\rho +p)n\Omega _{n}\tilde{r}%
_{A}^{n-1}d\tilde{r}_{A},
\end{eqnarray}
where $T_{m}$ is the temperature of the matter field inside the apparent
horizon. We assume the temperature of the perfect fluid inside the apparent
horizon scales as the temperature of the apparent horizon $T_{h}$. Thus, we
suppose the temperature $T_{m}=T_{h}$. We limit ourselves to the assumption
of the local equilibrium hypothesis, that the energy would not spontaneously
flow between the horizon and the fluid, the latter would be at variance with
the FRW geometry. Therefore, from the Gibbs equation, we get
\begin{equation}
T_{m}\dot{S}_{m}=n\Omega _{n}{\tilde{r}}_{A}^{n-1}(\rho +p)\dot{\tilde{r}}%
_{A}-n\Omega _{n}{\tilde{r}}_{A}^{n}H(\rho +p), \label{TdSm}
\end{equation}
where we have used the continuity equation (\ref{Cont}). Adding equations (%
\ref{TSh2}) and (\ref{TdSm}), after substituting $\dot{\tilde{r}}_{A}$ from
Eq. (\ref{dotr1}), we obtain
\begin{equation*}
T_{h}(\dot{S}_{h}+\dot{S}_{m})=\frac{4\pi G_{n+1}}{n-1}AH(\rho +p)^{2}{%
\tilde{r}}_{A}^{3}\times \left[ 1+\frac{2\hat{\mu}_{2}l^{2}}{\tilde{r}%
_{A}^{2}}+\frac{3\hat{\mu}_{3}l^{4}}{\tilde{r}_{A}^{4}}+\frac{4\hat{\mu}%
_{4}l^{6}}{\tilde{r}_{A}^{6}}+...\right] ^{-1}.
\end{equation*}
Expanding the r.h.s. of the above equation for the late time where $\tilde{r}%
_{A}\gg l$, we arrive at
\begin{equation*}
T_{h}(\dot{S}_{h}+\dot{S}_{m})=\frac{4\pi G_{n+1}}{n-1}AH(\rho +p)^{2}{%
\tilde{r}}_{A}^{3}\times \left[ 1-\frac{2\hat{\mu}_{2}l^{2}}{\tilde{r}%
_{A}^{2}}-\frac{3\hat{\mu}_{3}l^{4}}{\tilde{r}_{A}^{4}}-\frac{4\hat{\mu}%
_{4}l^{6}}{\tilde{r}_{A}^{6}}+...\right] .
\end{equation*}
The expression in the bracket, in the late time cosmology is positive which
indicate that $\dot{S}_{h}+\dot{S}_{m}\geq 0$. This implies that for the
late time cosmology, the GSL of thermodynamics is fulfilled in the universe
governed by quasi-topological gravity, regardless of the nature of the
energy content of the universe.

\section{Conclusions\label{Con}}

In this paper we investigated the thermodynamical properties of the apparent
horizon in quasi-topological gravity. We first derived the Friedmann
equation governing the evolution of the universe in quartic and higher order
quasi-topological gravity by varying the corresponding action of
quasi-topological gravity. Then, by applying the first law of
thermodynamics, $dE=T_{h}dS_{h}+WdV$ on the apparent horizon of FRW
universe, we extracted the Friedmann equation of quartic and higher order
quasi-topological gravity. Here $E=\rho V$ is the total energy inside the
apparent horizon and $T_{h}$ and $S_{h}$ are the temperature and entropy
associated with the apparent horizon, respectively. The entropy expression
depends on the gravity theory and it is generally accepted that the apparent
horizon entropy in each gravity theory has the same expression as the
entropy of black hole horizon, but replacing the black hole horizon radius $%
r_{+}$ with the apparent horizon radius ${\tilde{r}_{A}}$. We found that the
result obtained by the first law is exactly coincides with the ones derived
from the action principle. We also studied the time evolution of the total
entropy, including the entropy associated with the apparent horizon together
with the matter field entropy enveloped by the apparent horizon. Our study
implies that, with the assumption of the local equilibrium hypothesis, the
GSL of thermodynamics is preserved for the late time cosmology in the
universe governed by quasi-topological gravity.

\acknowledgments{We are grateful to the referee for constructive comments which helped us to improve the paper significantly. We also thank Shiraz University Research Council. This work has been supported financially by Research Institute for Astronomy and Astrophysics of Maragha (RIAAM), Iran.}

\end{document}